\documentclass{ws-mpla}
\usepackage[super]{cite}
\usepackage{graphicx}
\begin{document}

\markboth{X. Calmet and M. Keller}
{Cosmological Evolution of Fundamental Constants}

\catchline{}{}{}{}{}

\title{Cosmological Evolution of Fundamental Constants: From Theory to Experiment}

\author{\footnotesize Xavier Calmet$^a$ and Matthias Keller$^b$}

\address{
Physics $\&$ Astronomy, 
University of Sussex,   Falmer, Brighton, BN1 9QH, United Kingdom 
$^a$x.calmet@sussex.ac.uk, $^b$m.k.keller@sussex.ac.uk}

\maketitle

\pub{Received (Day Month Year)}{Revised (Day Month Year)}

\begin{abstract}
In this paper we discuss a possible cosmological time evolution of fundamental constants from the theoretical and experimental point of views. On the theoretical side, we explain that such a cosmological time evolution is actually something very natural which can be described by mechanisms similar to those used to explain cosmic inflation. We then discuss implications for grand unified theories, showing that the unification condition of the gauge coupling could evolve with cosmological time. Measurements of the electron-to-proton mass ratio can test grand unified theories using low energy data. Following the theoretical discussion, we review the current status of precision measurements of fundamental constants and their potential cosmological time dependence. 

\end{abstract}

\ccode{PACS Nos.: 06.20.Jr,04.20.Cv,98.80.Cq}

\section{Introduction}

The interest in a the cosmological time evolution of physical laws or of constants of nature has a long history dating back to Dirac's large number hypothesis \cite{dirac1,dirac2}. More recently, there has been conflicting information coming from astrophysical observations using distant quasars: the Keck/HIRES group \cite{Webb:2000mn} claims to have discovered a time variation of the fine structure constant on the level of 5$\sigma$, while the VLT/UVES group \cite{Chand:2004ct} claims to see no sign of such a time variation, see e.g. \cite{Uzan:2010pm} for a recent review. Both observations are using quasars at similar distances from Earth.

Given the potentially conflicting observational situation laboratory based experiments may provide further information about the constancy of physical constants. In this article we explain how a cosmological evolution of fundamental constants can occur without the need for a violation of Lorentz invariance or violations of the equivalence principle. We will show that whenever a scalar field roles down a potential, and when this scalar field vacuum expectation value fixes some of the physical parameter, one can expect some cosmological evolution of these parameters. In that sense, a time variation of physical constants is not something unexpected. It can be explained with very well understood physics.  We shall then discuss how a cosmological evolution of physical constants could probe grand unified theories using very low energy physics. Then we describe experiments designed to test the time variation of the fine structure constant or of the ratio of the electron mass to the proton mass.

\section{Theoretical motivations}
\subsection{Cosmological evolution of physical constants}

Fundamental theories of physics are based on the notion of symmetries which enable one to classify particles according to the representations of the Lorentz group and their interactions according to representations of gauge groups. While symmetries are rather restrictive and constraining on the type of interactions between the particles which can be introduced in the model, the strength of these interactions and the masses of such particles are free parameters. These leads to a plethora of fundamental constants. Within the standard model of particle physics, there are 28 fundamental parameters (note that 22 of these parameters are needed to describe fermion masses). In this number we have not counted the speed of light $c$ or Planck's constant $\hbar$. The number of fundamental constants is even larger if we include other cosmological parameters such as the cosmological constant or the non-minimal coupling of the Higgs boson to the Ricci scalar.

Modern theories of nature are based on renormalizable quantum field theories. Within this mathematical framework, it is impossible to calculate the value of the coupling constants from first principles.  There is one class of models \cite{tHooft:2011aa} with no free parameters, but these models are an exception and they are far from describing the real world. Within typical gauge theories, coupling constants are renormalized and depend on the energy scale at which they are probed. Because quantum gravity does not seem to be described by a renormalizable quantum field theory, extensions such as non-commutative geometry \cite{Connes:1994yd} or string theory \cite{Polchinski:1998rq,Polchinski:1998rr} have been considered. While it is difficult to make the link between these models and experiments, they are interesting because because within these models some, if not all, fundamental constants are in principle calculable. In  particular, in string theory,  coupling constants are fixed by the expectation values of moduli, which are scalar fields, and are thus, at least in theory, calculable.

When a coupling constant is fixed by the expectation value of a scalar field, it is natural to expect a cosmological evolution of the value of this parameter. This can be best illustrated by considering the model of Higgs inflation \cite{Bezrukov:2007ep,Barvinsky:2009fy}. In the Einstein frame, the part of the action which describes  the scalar field sector of the model and the general relativity is given by
\begin{eqnarray}
S=-\int d^4 x \sqrt{-g} \left [ \frac{1}{2} M_P^2 R -\frac{1}{2} \partial_\mu \chi \partial^\mu \chi + U(\chi) \right ]
\end{eqnarray}
where $\chi$ is the Higgs field in the Einstein frame. The scalar potential $U(\chi)$ is given
\begin{eqnarray}
U(\chi)= \frac{1}{\Omega(\chi)^2} \frac{\lambda}{4} ( h(\chi)^2 -v^2)^2
\end{eqnarray}
with
\begin{eqnarray}
\Omega(\chi)^2=1+\xi \frac{h(\chi)^2}{M_P}
\end{eqnarray}
and
\begin{eqnarray}
\frac{d\xi}{d h}=\sqrt{\frac{\Omega^2 + 6 \xi^2 h^2/M_P^2}{\Omega^4}}.
\end{eqnarray}
In this model one assumes that the non-minimal coupling $\xi$ of the Higgs boson to the Ricci scalar is large. For small values of $h$ (equivalently of $\chi$), $U(\chi)$ approaches the usual Higgs potential. However, for large values of $h\gg M_P/\sqrt{\xi}$, the potential is exponentially flat. In the Higgs inflation model, one assume that the universe starts with a large Higgs field which rolls down the exponentially flat potential, thereby inflating our universe, before settling down at the minimum of the potential at its current value $v=246$ GeV.

It is interesting to note that the masses of the fermions and gauge bosons 
\begin{eqnarray}
m_{f,V}(\chi) = \frac{m(v)}{v} \frac{h(\chi)}{\Omega(\chi)}
\end{eqnarray}
are dependent of the value of the background Higgs field while the QCD term in the Lagrangian remains invariant
\begin{eqnarray}
-\frac{1}{4} G_{\mu\nu}^a G^{a \mu\nu}.
\end{eqnarray}
This implies that the electroweak masses scale changes with the cosmological evolution of the Higgs boson while the QCD mass scale is independent of the cosmological evolution. The cosmological evolution of the Higgs field would lead to a cosmological time dependence of the ratio
\begin{eqnarray}
\mu = \frac{m_e}{m_{p}}
\end{eqnarray}
where $m_e$ is the electron mass and $m_{p}$ is the proton mass which is essentially determined by the QCD scale. This model illustrates how dimensionless ratios of scales can have a cosmological evolution.  Also a ``spatial" dependence of a fundamental coupling constant could be due to the evolution of a scalar field in a potential in some parts of the universe while the same scalar field has already reached the minimum of the potential in another part. Clearly, inflation is a phenomenon which took place very early in the history of our universe and it ended very soon after the big bang. But it is conceivable that some moduli, not responsible for inflation, are still evolving and thereby leading to a cosmological evolution of some of the parameters of the standard model today which could be observable in laboratory based experiments.

Let us stress that it is crucial to differentiate cosmological evolution from time dependence. Coupling constants must be Lorentz scalars. There are very tight bounds on Lorentz violation, see e.g. \cite{Kostelecky:2003fs,Kostelecky:2008ts}. However, a cosmological evolution does not imply Lorentz violation, as shown in the example above. Another example is the model called quintessence, see e.g. \cite{Dvali:2001dd}.

Within the standard model of particles, there are many quantities which could have a cosmological evolution but it is difficult predict the variation of the fundamental constants. However, there is a class of models which predicts relations between some of the fundamental constants. These grand unified theories could be tested by measurements of the cosmological evolution of some of its parameters.

\subsection{Grand unification and cosmological time evolution of physical parameters}

Grand unified theories \cite{Fritzsch:1974nn,GeorgiGlashow} are very attractive as they tend to reduce the number of fundamental constants. We will show that a cosmological evolution of a Higgs boson multiplet in a grand unified theory could lead to a modification of the unification condition for the gauge couplings of the model. Measurement of a cosmological evolution of the electron-to-proton mass ratio would probe the nature of the grand unified theory.

\subsubsection{Cosmological time evolution of the unification condition}
LEP data indicates that the gauge couplings of the standard model do not meet into one point at the unification scale \cite{Amaldi:1991cn}. Could our universe have started from a more symmetrical state in which all gauge couplings unified at the unification scale? Let us consider operators of the type
\begin{eqnarray} \label{dim5}
\frac{c}{\bar M_P}\mbox{Tr} [\Phi G_{\mu\nu} G^{\mu\nu}]
\end{eqnarray}
where $\bar M_P$ is the reduced Planck mass ($2.43\times 10^{18}\,{\rm GeV}$), $c$ is a Wilson coefficient of order unity, $G_{\mu\nu}$ is the grand unified theory field strength and $\Phi$ is a scalar multiplet. The effects of such operators have been considered before \cite{Hill:1983xh,Shafi,Hall:1992kq,Vayonakis:1993nn,Rizzo:1984mk,Datta:1995as,Dasgupta:1995js,Huitu:1999eh,Tobe:2003yj,Chakrabortty:2008zk,Calmet:2008df,Calmet:2008er,Calmet:2009hp}, where it was shown that these operators lead to a modification of the unification condition. Here we propose that the unification condition could have a cosmological time evolution. 

Let's consider SU(5) grand unification with the multiplet $\Phi$ in the adjoint representation. The adjoint Higgs field has a vacuum expectation value $\left\langle H \right\rangle = M_X \left(2,2,2,-3,-3\right)/\sqrt{50\pi\alpha_G}$, where $M_X$ is the unification scale, $\alpha_G$ is the value of the SU(5) gauge coupling at the unification scale. The operator (\ref{dim5}) modifies the kinetic terms of the SU(3)$\times$SU(2)$\times$U(1)  gauge bosons:
\begin{equation}
\label{gaugekineticterm}
\begin{split}
-&\frac{1}{4} \left(1+\epsilon_1\right)F_{\mu\nu} F^{\mu\nu}_{{\rm U}(1)}
-\frac{1}{2}\left(1+\epsilon_2\right){\rm Tr}\left(F_{\mu\nu} F^{\mu\nu}_{{\rm SU}(2)}\right)\\
& -\frac{1}{2}\left(1+\epsilon_3\right){\rm Tr}\left(F_{\mu\nu} F^{\mu\nu}_{{\rm SU}(3)}\right)
\end{split}
\end{equation}
with
\begin{equation}
\label{epsilons}
\epsilon_1=\frac{\epsilon_2}{3}=-\frac{\epsilon_3}{2}=\frac{\sqrt{2}}{5\sqrt{\pi}}\frac{c}{\sqrt{\alpha_G}}\frac{M_X}{M_{Pl}}.
\end{equation}
A finite field redefinition $A_{\mu}^{i} \to \left(1+\epsilon_i\right)^{1/2} A_{\mu}^{i}$ leads to the familiar form for the kinetic terms. The corresponding redefined coupling constants $g_i \to \left(1+\epsilon_i\right)^{-1/2} g_i$ are observed at low energies. In terms of the observable rescaled couplings, the unification condition reads:
\begin{equation}
\label{boundarycondition}
\alpha_G = \left(1+\epsilon_1\right) \alpha_1(M_X)=\left(1+\epsilon_2\right) \alpha_2(M_X) \\
 = \left(1+\epsilon_3\right) \alpha_3(M_X)~.
\end{equation}
Let's now imagine that the adjoint Higgs roles down a potential in analogy to the Higgs inflation model discussed above. Effectively this account for $M_X$ to have a cosmological evolution and we can see immediately how the unification condition could change with time. Depending on the value of the Wilson coefficient, we could have had a numerical unification of the gauge coupling at the birth of universe which is now spoiled by the evolution of $M_X$.

\subsubsection{A Time Variation of Proton-Electron Mass Ratio}

Measurements of the proton-electron mass ratio could be used to probe grand unified theories\cite{Calmet:2001nu,Calmet:2002ja,Calmet:2002jz,Langacker:2001td,Campbell:1994bf,Olive:2002tz,Dent:2001ga,Dent:2003dk,Landau:2000cc,Wetterich:2003jt,Flambaum:2006ip}. We will work at the one loop level and ignore possible cosmological time variations of Yukawa couplings and of Higgs boson masses.  Within these approximations, we only have two parameters: the unification scale $M_X$ and  the unified coupling constant $\alpha_G$. The proton mass is determined mainly by the QCD scale and thus quark masses can be neglected.
We  focus on the QCD scale $\Lambda_{QCD}$ and extract its value from the Landau pole of the renormalization group equations for the couplings of the supersymmetric standard model:
\begin{eqnarray} \label{running}
  \alpha_3(\mu)^{-1}
 &=&
\frac{1}{\alpha_3(\Lambda_I)}+\frac{1}{2 \pi}
  b_3   \ln
  \left ( \frac{M_X}{\mu} \right) 
  \end{eqnarray}
where the parameters $b_i$ are given by $b_i\!\!\!=(b_1,
b_2, b_3)=(41/10, -19/6, -7)$. The QCD scale is given by
\begin{eqnarray}
\Lambda_{QCD}= M_X \exp\left(\frac{2 \pi}{\alpha_G}\right)^{\frac{1}{b_3}}.
\end{eqnarray}
The time variation of $\Lambda_{QCD}$ is then determined by
\begin{eqnarray}
\frac{\dot \Lambda_{QCD}}{\Lambda_{QCD}}= -\frac{2\pi}{b_3} \frac{\dot \alpha_G}{\alpha_G^2}
+ \frac{\dot M_X}{M_X}.
\end{eqnarray}
For a constant electron mass this equation determines the ratio $\frac{\Delta \mu}{\mu}$:
\begin{eqnarray} \label{result}
\frac{\Delta \mu}{\mu}= -\frac{2\pi}{b_3} \frac{\dot \alpha_G}{\alpha_G^2}
+\frac{\dot M_X}{M_X} = \frac{2 \pi}{7}  \frac{\dot \alpha_G}{\alpha_G^2} + \frac{\dot M_X}{M_X}.
\end{eqnarray}
Note that a measurement of the time variation of the electron-to-proton mass ratio provides a direct determination of the time dependence of the unified coupling constant:
\begin{eqnarray}
\frac{\dot \alpha_G}{\alpha_G^2}=\frac{3}{8}\frac{\dot \alpha}{\alpha^2}
\end{eqnarray}
since this relation is renormalization scale invariant. 

\section{Experimental Status}

As described in the previous sections, there are compelling theoretical arguments for a cosmological evolution of fundamental constants. In addition, spatial variations seem to be possible. In order to be able to measure these possible changes, it must be possible to relate these constants to measurable quantities. In addition, due to the expected small scale of these potential changes, the accuracy of these quantities must be very high. 
By measuring these quantities over time, an evolution of fundamental constants can be detected. Currently, there are two approaches to probe temporal variations. Firstly, fundamental constants can be deduced from observations or measurements of naturally occurring phenomena in the distant past. While these type of measurements often suffer from the lack of knowledge of the environmental conditions, they have a significant role as they provide access to the values of fundamental constants over long time scales.

One of the first constraints on the fine structure constants comes from this type of approach: analyzing the abundance of specific isotopes in a naturally occurring fission reactors, the Oklo phenomenon (e.g.\cite{Shlyakhter, Petrov}). The measured variations of $-0.9\cdot10^{-7}<\frac{\Delta \alpha}{\alpha}<1.2\cdot 10^{-7}\;$\cite{Shlyakhter} for the early data analysis and $-0.24\cdot10^{-7}<\frac{\Delta \alpha}{\alpha}<0.11\cdot 10^{-7}\;$\cite{Gould} for a more recent analysis are consistent with no variation of the fine structure constant over an time of about $1.8\cdot 10^9$ years.
In order to deduce $\alpha$ from the measured isotope abundances, several assumptions have to be made such as the details of the geometry, the neutron spectrum or the quality of the extracted samples. These contribute to the error of the measured variations. 
Similarly, by investigating the decay of radio-isotopes in meteorites, Dyson\cite{Dyson} measured a variation in $\alpha$ of $\left|\frac{\Delta\alpha}{\alpha}\right|<4\cdot 10^{-4}$ which corresponds to a limit of the linear change of $\left|\frac{\dot{\alpha}}{\alpha}\right|< 2\cdot 10^{-13}$yr$^{-1}$. 
Both methods for deducing the fine structure constant from the measured data rely on a set of assumptions and on the model of the nuclear phenomena. The currently best constrains on the variations of fundamental constants over long time scales come form astronomical observations which will be discussed in more detail below.\\
The second approach to test temporal variations of fundamental constants is to employ laboratory based precision measurements. Due to the unparalleled control over the environmental conditions, laboratory based measurements are ideal tools to determine the current value of fundamental constants. By measuring over a longer time period, their changes can be probed. However, due to the limited accessible time interval, the detectable rate of change is limited. Nevertheless, laboratory based measurement provide important information about possible variation in our epoch.\\ 

The currently most stringent constrains on variations of fundamental constants relies their effect on the atomic and molecular structure. While effects of the weak force haven't been demonstrated with such systems due to the tiny impact, the electromagnetic and strong interactions have profound impacts on atomic and molecular systems. Thus the internal energy level structure and therefor the atomic and molecular transitions are prime candidates for measuring changes of fundamental constants. 
In order to characterize the sensitivity of atomic or molecular transitions with respect to the changes $\Delta \alpha$ and $\Delta \mu$ in $\alpha$ and $\mu$, relative sensitivity parameters $K_{\alpha}$ and $K_{\mu}$ are introduces as follows:
\begin{equation}
\frac{\Delta f}{f} = K_{\alpha}\frac{\Delta \alpha}{\alpha}+K_{\mu}\frac{\Delta \mu}{\mu},
\end{equation}
with the fractional frequency variation $\frac{\Delta f}{f}$ of a given transition.
To detect variation in $\alpha$ or $\mu$ the frequency of transitions with different sensitivities must be employed so that a frequency change can be measured. However, for spectroscopy the absolute sensitivity q of an atomic transition is often more important. It is related to the relative sensitivity through:
\begin{equation}
q_{\alpha,\mu}=\frac{1}{2} K_{\alpha,\mu} f,
\end{equation}
with the transition frequency $f$.\\

\subsection{Fine structure constant}

The fine structure constant determines directly the energies of electronic states of atoms and molecules with the electronic transition frequency of the gross structure scaling with the Rydberg constant $R_\infty$, the fine structure with $R_\infty\cdot Z^2 \cdot \alpha^2$ and the hyperfine structure with $R_\infty\cdot Z^2 \cdot \alpha^2 \cdot g_i \cdot \mu$. Here Z denotes the nuclear charge and $g_i$ the nuclear gyromagnetic factor. In general, electronic fine structure transition frequencies can be approximated by $\nu_{\mbox{\scriptsize fs}} \propto R_\infty \cdot F_{\mbox{\scriptsize fs}}(\alpha)$ \cite{Karshenboim} in which $F_{\mbox{\scriptsize fs}}(\alpha)$ is a dimensionless function which takes the internal structure into account. Similarly, hyperfine transitions can be expressed as $\nu_{\mbox{\scriptsize hfs}} \propto R_\infty \cdot F_{\mbox{\scriptsize hfs}}(\alpha)\cdot g_i \cdot \mu$ with the dimensionless function $F_{\mbox{\scriptsize hfs}}(\alpha)$. The functions $F_{\mbox{\scriptsize fs}}(\alpha)$ and $F_{\mbox{\scriptsize hfs}}(\alpha)$ depend on the details of the atomic structure and have been calculated ab inito for a range of transitions (e.g. \cite{Dzuba, Dzuba2, Dzuba3, Dzuba4, Flambaum}). Some transitions such as the hydrogen 1s-2s transition and the $^1S_{0}\rightarrow ^3P_{0}$ transition in $^{27}$Al$^+$ have a very small $\alpha$ dependence and can serve as anchors for reference measurements. Other transitions exhibit large sensitivities. For example, rare earth atoms and ions have been shown to posses relative sensitivities of up to 100,000 \cite{Angstmann}.
Similarly, highly charged heavy atomic ions are highly sensitive to changes in $\alpha$ \cite{Berengut2}, due to the dependence of the sensitivity on the nuclear charge and the ionization potential. With an absolute sensitivity coefficient of 355,000 cm$^{-1}$ for the $5f^2(J=4) \rightarrow 5f6p(J=3)$ transition Cf$^{16+}$ has been suggested to have the largest known sensitivity. 
Unfortunately, these systems are very hard to work with in an laboratory experiment due to their large charge and complex level structure. Furthermore, they are unavailable for astronomic observations. Comparing highly sensitive transitions with anchor transitions over extended time periods provide a measurement of variation of fundamental constants.

Employing spectroscopic methods to detect possible variations of fundamental constants gives access not only to ultra-precise laboratory measurements but also to possible changes in constants in the distant past through astronomical observations.
Observing the absorption spectrum of interstellar clouds possible changes of the fine structure constant have been investigated. Taking advantage of a multitude of atomic transitions of different species, the environmental condition of the absorption medium can be probed with some anchor transitions such as in MgI and MgII. Measuring sensitive transitions within the same medium allows then to extract potential changes in $\alpha$. Observing the light emission from several quasars and looking at the absorption spectrum in interstellar clouds, values for possible variations in $\alpha$ at different red-shifts and thus look back times have been measured at the Keck telescope $\left.\frac{\Delta\alpha}{\alpha}\right|_{z<1.8}=(-0.54 \pm 0.12)\cdot 10^{-5}$ and $\left.\frac{\Delta\alpha}{\alpha}\right|_{z>1.8}=(-0.74 \pm 0.17)\cdot 10^{-5}\;$ \cite{Murphy}. Measurements with the VLT, however, contradict these observations with measured variations of $\left.\frac{\Delta\alpha}{\alpha}\right|_{z<1.8}=(-0.06 \pm 0.16)\cdot 10^{-5}\;$ \cite{Webb} and $\left.\frac{\Delta\alpha}{\alpha}\right|_{z>1.8}=(+0.61 \pm 0.20)\cdot 10^{-5}\;$ \cite{Srianand}. However, this discrepancy may be a result of a spatial variation of $\alpha$ \cite{Webb}.
Alternatively to employing multiple species, different transitions within one atomic species can be employed. For instance, one transition with a low $\alpha$ sensitivity is employed to probe the environmental condition while the fine structure constant is measured with another transition. In this way several systematic uncertainties are reduced which results in a significant improvement of the measured variations.  Several observations have been performed resulting in $\left.\frac{\Delta\alpha}{\alpha}\right|_{z=1.15}=(-0.12 \pm 1.79)\cdot 10^{-6}$ \cite {Molaro}, $\left.\frac{\Delta\alpha}{\alpha}\right|_{z<1.15}=(0.5 \pm 2.4)\cdot 10^{-6 }\;$ \cite{Chand} and $\left.\frac{\Delta\alpha}{\alpha}\right|_{z=1.84}=(-5.66 \pm 2.67)\cdot 10^{-6}\;$ \cite{Levshakov, Molaro}.

Even though astronomical observations allow access to large look back times and remote locations of the universe, the errors due to the uncertainty in the environmental parameter of the absorption medium and systematic uncertainties limit the minimal detectable change in $\alpha$ to currently $\left.\frac{\Delta\alpha}{\alpha}\right|_{max}<\pm 1.6\cdot 10^{-6 }\;$ \cite{Webb}. With long look back times and the possibility to probe remote areas of the universe astronomical observations are best suited to detect remnants of the cosmological evolution of fundamental constants. Even though most of this evolutions is likely to be confined to the epoch of inflation, there may be some residual drift still present at accessible red shifts. Even today, fundamental constants may vary. 
Due to the unparalleled control over the environmental conditions, lab based high resolution spectroscopy of  atomic transitions are attractive systems to measure changes of fundamental constants. Event though the 'look back time' of these systems is limited to a few years, the achievable accuracy makes the detection of variations with laboratory based experiments feasible. In addition to temporal changes, spatial variations can be probes due to the earth's motion through the universe. Berengut \cite{Berengut3} et al., for example, predict that the earth is moving towards an area of higher $\alpha$. In the last years, there have been a large range of different laboratory experiments to probe variations in $\alpha$. 
Comparing hyperfine transitions in rubidium and cesium, S. Bize et al.\cite{Bize} have measured an upper bond of changes in $\left|\frac{g_{\mbox{\tiny Cs}}}{g_{\mbox{\tiny Rb}}}\alpha^{0.49}\right| < 5.3\cdot 10^{-16}$yr$^{-1}$.\\
By comparing transitions of various chemical elements with the cesium atomic clock, constrains on variations of fundamental constants can be improved. Due to the hyperfine reference transition of cesium, the frequency ratios are always codependent on the nuclear magnetic moment and the electron-to-proton mass ratio. 
Employing two different transitions within $^{171}$Yb$^+$ ions and comparing them with a cesium atomic clock, constrains on the electron-to-proton mass ratio as well as on the fine structure constant have been recently determined \cite{Godun, Peik}. The resulting variation $\frac{\dot{\alpha}}{\alpha}= (-0.7\pm 2.1)\cdot 10^{-17 }$yr$^{-1}$ and $\frac{\dot{\mu}}{\mu}= (-0.2\pm 1.1)\cdot 10^{-16}$yr$^{-1}$ which assume constant quark masses \cite{Godun} and $\frac{\dot{\alpha}}{\alpha}= (-0.20\pm 10)\cdot 10^{-16 }$yr$^{-1}$ and $\frac{\dot{\mu}}{\mu}= (-0.5\pm 1.6)\cdot 10^{-16}$yr$^{-1}$  which allows for potential changes in the quark masses \cite{Peik2} are currently the best limits from laboratory experiments.
Employing narrow optical transitions in Hg$^+$ and Al$^+$ ions directly allows to measure changes in $\alpha$ without co-dependencies. Comparing the transition frequencies over 12 month, drift rates of $\frac{\dot{\alpha}}{\alpha}= (-1.6\pm 2.3)\cdot 10^{-17 }$yr$^{-1}$ have been measured \cite{Rosenband}. \\

\subsection{Electron-to-proton mass ratio}
While atomic and molecular electronic gross and fine structure transitions are sensitive to variations of the fine structure constant, their dependence on the electron-to-proton mass ratio is comparably small. Typical relative sensitivity parameters are on the order of $10^{-2}$. By measuring frequency ratios between hyperfine and other atomic transitions, limits on the variation of $g_{\mbox{\tiny Cs}}\mu$ can be determined by taking other constrains of $\alpha$ variations into account.
Apart from hyperfine transitions in atoms, molecules offer another access to the electron-to-proton mass ratio through ro-vibrational transitions.
While the pure electronic transitions in molecules, similar to atoms, have a sensitivity coefficient $K_{\mu}$ on the order of 1\%, vibrational transitions and rotational transitions have a sensitivity of -1/2 and -1 respectively. Several molecular species have been proposed for measuring potential variations in $\mu$, e.g. \cite{Salumbides, Beloy, Nijs, Schiller, Keller}.\\
Exploiting near degeneracies between ro-vibrational and electronic levels the sensitivity can be significantly enhanced \cite{Flambaum3}. E.g. utilizing the near degeneracy of the molecular spin-orbit coupling and the molecular rotation, sensitivities as large as 460 have been predicted \cite{Kozlov}.
For polyatomic species large amplitude motions can exhibit a high sensitivity to variations in $\mu$. It has been shown that the tunneling inversion of the ammonia molecule have a relative sensitivity of -4.4 and -3.4 for the first and second inversion mode respectively. This sensitivity is significantly enhanced by exploiting near degeneracies between inversion and rotational levels in different isotopologues \cite{Veldhoven, Flambaum2}. 
Also internal rotations can exhibit significant enhancements of the sensitivity as predicted by \cite{Jansen2, Ilyushin, Jansen3}.\\
Employing transitions in $H_2$ astronomic observations with the Keck telescope and the VLT variations of $\frac{\Delta \mu}{\mu}=(5.6\pm 6.1 )\cdot 10^{-6}$ at z=2.059 \cite{Malec} and $\frac{\Delta \mu}{\mu}=(8.5\pm 4.2 )\cdot 10^{-6}\;$ \cite{Weerdenburg} respectively have been observed. Using the radio telescope in Effelsberg, Bagdonaite et al. employed transitions within methanol to obtain $\frac{\Delta \mu}{\mu}=(0.0\pm 1.0 )\cdot 10^{-7}$ at red-shifts of 0.89 \cite{Bagdonaite}. 
The radio frequency absorption spectrum of methanol and ammonia from several astronomical sources lead to a constraint of $\frac{\Delta \mu}{\mu} < 10^{-7}\;$ \cite{Jansen}.
Laboratory based limits on variations of the electron-to-proton mass ratio have been be derived from comparing the hyperfine clock transition of cesium with ro-vibrational transitions of SF$_6$ \cite{Shelkovnikov}. By combining the measured frequency ratios between these transitions and taking a limits on the variations of the relative frequencies of the hydrogen maser and the cesium atomic clock into account a limit on the pure $\mu$ variation of $\frac{\dot{\mu}}{\mu}= (-3.8\pm 5.6)\cdot 10^{-14 }\:$yr$^{-1}$ can be obtained. However, the currently most stringent constrain comes from the comparison of the transitions in Yb$^+$ \cite{Godun, Peik} with the cesium atomic clock, as discussed in the previous section.

\section{Conclusions}

We have shown that a cosmological evolution of fundamental constants is something very natural within our current understanding of physics. Mechanisms very similar to those advocated to explain the inflation of our universe could give rise to such a cosmological evolution. We have stressed that it is crucial to differentiate between time variation and cosmological evolution. The former would imply a violation of Lorentz invariance which is severely constrained by experiment, while the latter is something we naturally expect to take place when scalar fields role down a potential before settling down in its minimum. In that sense, a cosmological evolution is linked to the question of initial conditions of our universe: where does the scalar field sit at the time of the big bang. The flatness of the potential determines how long the scalar field will role before settling down. We have shown that a cosmological time evolution of the unification condition for the gauge couplings of a grand unified theory could naturally occur and explain why LEP did not measure a perfect numerical unification of the gauge couplings of the standard model. We then reviewed the well known fact, namely that a cosmological evolution of the electron-to-proton mass ratio would enable one to probe the nature of the grand unified theory by performing only low energy measurements.

On the experimental physics side, we have shown that to isolate the variation of a fundamental constant the results of several measurements have to be combined. Thus by comparing the transition frequencies of Al$^+$ and $^{199}$Hg$^+$ or the transition frequencies of the quadrupole and octupole transition within Yb$^+$ limits on pure $\alpha$ variations can be found. 
This shows that it is important to collect data from a multitude of different systems to extract the possible variation of specific fundamental constants. This is also helpful to eliminate effects of the specific models that are used to describe the dependence of transitions from particular constants.\\
Combining astronomical observations with laboratory measurements provides insight into the underlying mechanism of the potential change of fundamental constants. As pointed out in the first part of this article, there are compelling models which associate a change in some fundamental constants on cosmological time scales with the inflation phase of the universe. With long look back times astronomical observations are best suited to observe remnants this effect as well as possible regions in the universe where the fundamental constants may have other values. On the other hand laboratory experiments may, due to their high accuracy, be able to detect these effects even today if the scalar fields which fix the values of fundamental constants are still rolling down their potentials today. Furthermore, laboratory experiments  could detect changes due to other mechanisms, some of which predict oscillatory behaviors of the fundamental constants on short time scales see e.g.\cite{Arvanitaki} where time variation effects due to dilaton dark matter were discussed or \cite{Stadnik:2013raa,Derevianko:2013oaa} where topological defects were identified as a possible source of time variation of fundamental constants. Finally, an interesting potential correspondence between the time variation of the QCD scale and the vacuum energy density of an expanding universe has been discussed in \cite{Fritzsch:2012qc}. There are thus several plausible mechanisms which could lead to a time variation of fundamental constants and it is important to build experiments to probe their constancy as accurately as possible.


\begin{thebibliography}{0}

\bibitem{dirac1}
P.~A.~M.~Dirac, {\it Nature (London) }{\bf 139}, 323 (1937).

\bibitem{dirac2}
P.~A.~M.~Dirac, {\it Proc. Roy. Soc. London A} {\bf 165}, 198 (1938).

\bibitem{Webb:2000mn} 
  J.~K.~Webb, M.~T.~Murphy, V.~V.~Flambaum, V.~A.~Dzuba, J.~D.~Barrow, C.~W.~Churchill, J.~X.~Prochaska and A.~M.~Wolfe,
  {\it Phys.\ Rev.\ Lett.} {\bf 87}, 091301 (2001).
  
\bibitem{Chand:2004ct} 
  H.~Chand, R.~Srianand, P.~Petitjean and B.~Aracil, {\it Astron.\ Astrophys.}  {\bf 417}, 853 (2004).
  
\bibitem{Uzan:2010pm} 
  J.~P.~Uzan, {\it Living Rev.\ Rel.} {\bf 14}, 2 (2011).

\bibitem{tHooft:2011aa} 
  G.~'t Hooft, {\it Found.\ Phys.}  {\bf 41}, 1829 (2011).
  
\bibitem{Connes:1994yd} 
  A.~Connes,
  ISBN-9780121858605.

\bibitem{Polchinski:1998rq} 
  J.~Polchinski,
  ``String theory. Vol. 1: An introduction to the bosonic string,''
  Cambridge, UK: Univ. Pr. (1998) 402 p


\bibitem{Polchinski:1998rr} 
  J.~Polchinski,
  ``String theory. Vol. 2: Superstring theory and beyond,''
  Cambridge, UK: Univ. Pr. (1998) 531 p

\bibitem{Bezrukov:2007ep} 
  F.~L.~Bezrukov and M.~Shaposhnikov, {\it Phys.\ Lett.\ B} {\bf 659}, 703 (2008).
  
\bibitem{Barvinsky:2009fy} 
  A.~O.~Barvinsky, A.~Y.~Kamenshchik, C.~Kiefer, A.~A.~Starobinsky and C.~Steinwachs,
  {\it JCAP }{\bf 0912}, 003 (2009).
   
\bibitem{Kostelecky:2003fs} 
  V.~A.~Kostelecky, {\it Phys.\ Rev.\ D} {\bf 69}, 105009 (2004).
  
\bibitem{Kostelecky:2008ts} 
  V.~A.~Kostelecky and N.~Russell, {\it Rev.\ Mod.\ Phys.} {\bf 83}, 11 (2011).
  
\bibitem{Dvali:2001dd} 
  G.~R.~Dvali and M.~Zaldarriaga,
    Phys.\ Rev.\ Lett.\  {\bf 88}, 091303 (2002).
    
\bibitem{Fritzsch:1974nn} 
  H.~Fritzsch and P.~Minkowski, {\it Annals Phys.}  {\bf 93}, 193 (1975).
  
 \bibitem{GeorgiGlashow}
  H.~Georgi and S.~L.~Glashow, {\it Phys.~Rev.~Lett.}{\bf 32}, 438 (1974).
  
\bibitem{Amaldi:1991cn}
  U.~Amaldi, W.~de Boer and H.~Furstenau, {\it Phys.\ Lett.\  B} {\bf 260}, 447 (1991).

\bibitem{Hill:1983xh}
  C.~T.~Hill, {\it Phys.\ Lett.\  B} {\bf 135}, 47 (1984).
  
\bibitem{Shafi}
  Q.~Shafi and C.~Wetterich, {\it Phys.\ Rev.\ Lett.}  {\bf 52}, 875 (1984).
  
\bibitem{Hall:1992kq}
  L.~J.~Hall and U.~Sarid, {\it Phys.\ Rev.\ Lett.} {\bf 70}, 2673 (1993).
   
\bibitem{Vayonakis:1993nn}
  A.~Vayonakis, {\it Phys.\ Lett.\  B} {\bf 307}, 318 (1993).
 
\bibitem{Rizzo:1984mk}
  T.~G.~Rizzo, {\it Phys.\ Lett.\  B} {\bf 142}, 163 (1984).
  
  \bibitem{Datta:1995as}
  A.~Datta, S.~Pakvasa and U.~Sarkar, {\it Phys.\ Rev.\  D} {\bf 52}, 550 (1995).

\bibitem{Dasgupta:1995js}
  T.~Dasgupta, P.~Mamales and P.~Nath, {\it Phys.\ Rev.\  D} {\bf 52}, 5366 (1995).

\bibitem{Huitu:1999eh}
  K.~Huitu, Y.~Kawamura, T.~Kobayashi and K.~Puolamaki, {\it Phys.\ Lett. B} {\bf 468}, 111 (1999).

\bibitem{Tobe:2003yj}
  K.~Tobe and J.~D.~Wells, {\it Phys.\ Lett.\  B} {\bf 588}, 99 (2004).

\bibitem{Chakrabortty:2008zk}
  J.~Chakrabortty and A.~Raychaudhuri, {\it Phys.\ Lett.\ B }{\bf 673}, 57 (2009).

\bibitem{Calmet:2008df}
  X.~Calmet, S.~D.~H.~Hsu and D.~Reeb,  {\it Phys.\ Rev.\ Lett.\ } {\bf 101}, 171802 (2008).
	
\bibitem{Calmet:2008er}
  X.~Calmet, S.~D.~H.~Hsu and D.~Reeb, AIP Conf.\ Proc.\  {\bf 1078}, 432 (2009).
  
\bibitem{Calmet:2009hp}
  X.~Calmet, S.~D.~H.~Hsu and D.~Reeb, {\it Phys.\ Rev.\  D} {\bf 81}, 035007 (2010).
  
\bibitem{Calmet:2001nu}
  X.~Calmet and H.~Fritzsch, {\it Eur.\ Phys.\ J.\ C} {\bf 24}, 639 (2002).

\bibitem{Calmet:2002ja}
  X.~Calmet and H.~Fritzsch, {\it Phys.\ Lett.\ B} {\bf 540}, 173 (2002).

\bibitem{Calmet:2002jz}
  X.~Calmet and H.~Fritzsch,
  ``Grand unification and time variation of the gauge couplings,'' in Hamburg 2002, Supersymmetry and unification of fundamental interactions, vol. 2, pages1301-1306.
	
\bibitem{Langacker:2001td} 
  P.~Langacker, G.~Segre and M.~J.~Strassler, {\it Phys.\ Lett.\ B} {\bf 528}, 121 (2002).
  
\bibitem{Campbell:1994bf} 
  B.~A.~Campbell and K.~A.~Olive, {\it Phys.\ Lett.\ B} {\bf 345}, 429 (1995).
  
\bibitem{Olive:2002tz} 
  K.~A.~Olive, M.~Pospelov, Y.~Z.~Qian, A.~Coc, M.~Casse and E.~Vangioni-Flam,
  {\it Phys.\ Rev.\ D} {\bf 66}, 045022 (2002).

\bibitem{Dent:2001ga}
  T.~Dent and M.~Fairbairn, {\it Nucl. Phys. B} {\bf 653}, 256 (2003).

\bibitem{Dent:2003dk}
  T.~Dent, arXiv:hep-ph/0305026. 
    
\bibitem{Landau:2000cc}
  S.~J.~Landau and H.~Vucetich,  {\it Astrophys. J.} {\bf 570}, 463 (2002).

\bibitem{Wetterich:2003jt}
  C.~Wetterich, {\it Phys. Lett. B} {\bf 561}, 10 (2003).
  
\bibitem{Flambaum:2006ip}
  V.~V.~Flambaum and A.~F.~Tedesco, arXiv:nucl-th/0601050.
 
\bibitem{Shlyakhter}
A.~I.~Shlyakhter, {\it Nature (London)} {\bf 264}, 340 (1976).

\bibitem{Petrov}
Yu. V. Petrov, A.I. Nazarov, M.S. Onegin, V.Yu. Petrov, et al., {\it Phys. Rev. C} {\bf 74}, 064610 (2006).

\bibitem{Gould}
C.R. Gould, E.I. Sharapov, and S.K. Lamoreaux, {\it Phys. Rev. C} {\bf 74}, 024607 (2006).

\bibitem{Dyson}
F.J. Dyson, The fundamental constants and their time variation, in “Aspects of Quantum
Theory”, A. Salam and E.P. Wigner Eds., p. 213 (Cambridge University Press, 1972).

\bibitem{Chou}
C. W. Chou, D. B. Hume, J. C. J. Koelemeij, D. J. Wineland, and T. Rosenband, {\it Phys. Rev. Lett.} {\bf 104}, 070802 (2010).

\bibitem{Karshenboim}
S.G. Karshenboim, {\it Gen. Rel. Grav.} {\bf 38} (2006) 159.

\bibitem{Dzuba}
V.A. Dzuba, and V.V. Flambaum, {\it Phys. Rev. A} {\bf 61}, 034502 (2001).

\bibitem{Dzuba2}
V.A. Dzuba, V.V. Flambaum, and M.V. Marchenko, {\it Phys. Rev. A} {\bf 68}, 022506 (2003).

\bibitem{Dzuba3}
V.A. Dzuba, V.V. Flambaum, and J.K. Webb, {\it Phys. Rev. A} {\bf 59}, 230 (1999).

\bibitem{Flambaum}
V.V. Flambaum, Limits on temporal variation of fine structure constant, quark masses and
strong interaction from atomic clock experiments, in Laser Spectroscopy, P. Hannaford, et
al. Eds. (World Scientific, 2004) p. 47.

\bibitem{Dzuba4}
V. A. Dzuba A. Derevianko and V. V. Flambaum, {\it Phys. Rev. A} {\bf 86}, 054502 (2012)

\bibitem{Angstmann}
E J Angstmann, V A Dzuba, V V Flambaum, A Yu Nevsky and S G Karshenboim
{\it J. Phys. B: At. Mol. Opt. Phys.} {\bf 39}, 1937–1944 (2006).

\bibitem{Berengut2}
J. C. Berengut, V. A. Dzuba, V.V. Flambaum, and A. Ong, {\it Phys. Rev. Lett.} {\bf 109}, 070802 (2012).

\bibitem{Murphy}
M.T. Murphy, V. V. Flambaum, J.K. Webb, V.V. Dzuba, et al., {\it Lect. Notes Phys.} {\bf 648}, 131 (2004).

\bibitem{Webb}
J.K. Webb, J.A. King, M.T. Murphy, V.V. Flambaum, et al., arxiv:1008.3907.

\bibitem{Srianand}
R. Srianand, H. Chand, P. Petitjean, and B. Aracil, {\it Phys. Rev. Lett.} {\bf 99}, 239002 (2007).

\bibitem{Molaro}
P. Molaro, D. Reimers, I.I. Agafonova, S.A. Levshakov, Bounds on the fine structure constant variability from FeII absorption lines in QSO spectra, in Atomic Clocks and Fundamental Constants, S. Karshenboim and E. Peik Eds.

\bibitem{Chand}
H. Chand, P. Petitjean, R. Srianand, {\it Aracil, Astron. Astrophys.} {\bf 451}, 45 (2006).

\bibitem{Levshakov}
S.A. Levshakov, P. Molaro, S. Lopez, S. D’Odorico, et al., {\it Astron. Astrophys.} {\bf 466}, 1077 (2007).

\bibitem{Berengut3}
J. C. Beregut and V. V. Flambaum, {\it Europhys. Lett.} {\bf 97}, 20006 (2012).

\bibitem{Bize}
S. Bize, P. Laurent, M. Abgrall, H. Marion, et al., {\it J. Phys. B}  {\bf 38}, S449 (2005).

\bibitem{Peik}
E. Peik, B. Lipphardt, H. Schnatz, T. Schneider, et al., {\it Phys. Rev. Lett.} {\bf 93}, 170801 (2004).

\bibitem{Godun}
R. M. Godun, P. B. R. Nisbet-Jones, J. M. Jones, S. A. King, L. A. M. Johnson, H. S. Margolis, K. Szymaniec, S. N. Lea, K. Bongs, P. Gill, 
{\it Phys. Rev. Lett} {\bf 113}, 210801 (2014).

\bibitem{Peik2}
N. Huntemann, B. Lipphardt, Chr. Tamm, V. Gerginov, S. Weyers, and E. Peik,
{\it Phys. Rev. Lett} {\bf 113}, 210802 (2014).

\bibitem{Rosenband}
[434] T. Rosenband, et al., {\it Science} {\bf 39}, 1808 (2008).

\bibitem{Salumbides}
E. J. Salumbides, M. L. Niu, J. Bagdonaite, N. de Oliveira, D. Joyeux, L. Nahon, and W. Ubachs, {\it Phys. Rev. A} {\bf 86}, 022510 (2012).

\bibitem{Beloy}
K. Beloy, M. G. Kozlov, A. Borschevsky, A. W. Hauser, V. V. Flambaum, and P. Schwerdtfeger, {\it Phys. Rev. A} {\bf 83}, 062514 (2011).

\bibitem{Nijs}
A. J. de Nijs, W. Ubachs, and H. L. Bethlem, {\it Phys. Rev. A} {\bf 86}, 032501 (2012).

\bibitem{Schiller}
S. Schiller and V. Korobov, {\it Phys. Rev. A} {\bf 71}, 032505 (2005).

\bibitem{Keller}
M. Kajita, G. Gopakumar, M. Abe, M. Hada and M. Keller, {\it Phys. Rev. A} {\bf 89}, 032509 (2014).

\bibitem{Flambaum3}
V. V. Flambaum and M. G. Kozlov, {\it Phys. Rev. Lett.}{\bf 99}, 150801 (2007).

\bibitem{Flambaum4}
V. V. Flambaum, Y. V. Stadnik, M. G. Kozlov, and A. N. Petrov, {\it Phys. Rev. A} {\bf 88}, 052124 (2013).

\bibitem{Kozlov}
M. G. Kozlov, {\it Phys. Rev. A} {\bf 80}, 022118 (2009).

\bibitem{Veldhoven}
J. van Veldhoven, J. Küpper, H. L. Bethlem, B. Sartakov, A. J. A. van Roij,
and G. Meijer, {\it Eur. Phys. J. D}{\bf 31}, 337 (2004).

\bibitem{Flambaum2}
V. V. Flambaum and M. G. Kozlov, {\it Phys. Rev. Lett.} {\bf 98}, 240801 (2007).

\bibitem{Jansen2}
P. Jansen, L. Xu, I. Kleiner, H.L. Bethlem, and W. Ubachs {\it Phys. Rev. A.} {\bf 87}, 052509 (2013).

\bibitem{Ilyushin}
V.V. Ilyushin, P Jansen, M.G. Kozlov, S.A. Levshakov, I. Kleiner, W. Ubachs, and H.L. Bethlem {\it Phys. Rev. A.} {\bf 85}, 032505 (2012).

\bibitem{Jansen3}
P. Jansen, I. Kleiner, L. Xu, W Ubachs, and H.L. Bethlem, {\it Phys. Rev. A.} {\bf 84}, 062505 (2011).

\bibitem{Malec}
A. L. Malec, R. Buning, M. T. Murphy, N.Milutinovic, S. L. Ellison, J. X. Prochaska, L. Kaper, J. Tumlinson, R. F. Carswell, and W. Ubachs, {\it Mon. Not. R. Astron Soc.} {\bf 403}, 1541 (2010).

\bibitem{Weerdenburg}
F. van Weerdenburg, M. T. Murphy, A. L. Malec, L. Kaper, and W.Ubachs, {\it Phys. Rev. Lett.} {\bf 106}, 180802 (2011).

\bibitem{Bagdonaite}
J. Bagdonaite, P. Jansen, C. Henkel, H. L. Bethlem, K. M. Menten, and W. Ubachs, {\it Science} {\bf 339}, 46 (2013).

\bibitem{Shelkovnikov}
A. Shelkovnikov, R. J. Butcher, C. Chardonnet, and A. Amy-Klein, {\it Phys. Rev. Lett.} {\bf 100}, 150801 (2008).

\bibitem{Jansen}
P. Jansen, H.L. Bethlem, and W. Ubachs, {\it J. Chem. Phys.} {\bf 140}, 010901 (2014).

\bibitem{Arvanitaki}
A. Arvanitaki, J Huang, and K. Van Tilburg, arXiv:1405.2925 (2014).


\bibitem{Stadnik:2013raa} 
  Y.~V.~Stadnik and V.~V.~Flambaum, Phys.\ Rev.\ D {\bf 89}, no. 4, 043522 (2014).
  

\bibitem{Derevianko:2013oaa} 
  A.~Derevianko and M.~Pospelov,  arXiv:1311.1244 [physics.atom-ph].
    
  
\bibitem{Fritzsch:2012qc} 
  H.~Fritzsch and J.~Sola,
  Class.\ Quant.\ Grav.\  {\bf 29}, 215002 (2012).

 
\end{thebibliography}
\end{document}